\documentclass[a4paper,keeplastbox]{jacow}
\begin{document}
\title{Machine detector interface for the $\rm e^{+}e^{-}$ future circular collider}
\author{M.~Boscolo\thanks{manuela.boscolo@lnf.infn.it}, O.R.~Blanco-Garcia, INFN/LNF, Frascati, Italy \\
N. Bacchetta\textsuperscript{1}, E.~Belli\textsuperscript{2}, M.~Benedikt, H.~Burkhardt, M.~Gil Costa,  
 K.~Elsener, \\E.~Leogrande, P.~Janot, H.~Ten Kate,
D.~El Khechen, A. Kolano,  R.~Kersevan, \\ M.~Lueckhof,  K. Oide,  E.~Perez, N.A. Teherani, O. Viazlo,
  Y. Voutsinas and \\ F. Zimmermann, CERN, Geneva, Switzerland\\
M.~Dam, Niels Bohr Institute, Copenhagen, Denmark\\
A.~Blondel, M. Koratzinos, DPNC/Geneva University, Geneva, Switzerland\\
A.~Novokhatski, M.~Sullivan, SLAC, Menlo Park, California, USA\\
A.~V.~Bogomyagkov, E.~B.~Levichev, S.~Sinyatkin, BINP SB RAS, Novosibirsk, Russia\\
F.~Collamati, INFN-Rome1, Rome, Italy\\
\textsuperscript{1}also INFN-Padova, Padova, Italy\\
\textsuperscript{2}also at University of Rome Sapienza and INFN-Roma1, Rome, Italy}

\maketitle
\begin{abstract}
The international Future Circular Collider (FCC) study~\cite{fccweb} aims at a design of p-p, $\rm e^{+}e^{-}$, e-p colliders to be built in a new 100~km tunnel in the Geneva region. 
The $\rm e^{+}e^{-}$ collider (FCC-ee) has a centre of mass energy range between 90 (Z-pole) and 375~GeV ($t\bar{t}$). To reach such unprecedented energies and luminosities,
   the design of the interaction region is  crucial. The  crab-waist collision scheme~\cite{ref:cw} has been chosen for the design and it will be compatible with all beam energies.
In this paper we will  describe the machine detector interface layout including the solenoid compensation scheme.  We will describe  how this layout fulfills all the requirements set by the parameters table and by the physical constraints. 
 We will summarize  the studies of the impact of the synchrotron radiation, the analysis of trapped modes and
of the  backgrounds induced by single beam and  luminosity effects giving an estimate of the losses in the interaction region and in the detector.
\end{abstract}
\section{\label{sec:1}Layout and design criteria}
The FCC-ee collider with 100~km circumference and a wide range of beam energies, from 45.6 to 182.5 GeV, aims at 
unprecedented levels of energies and luminosities. 
The requirements at the collision point for the accelerator and detector make the interaction region (IR) one of the most challenging parts of the overall design,
 this region is named machine detector interface (MDI).
 Table~\ref{tab:table1} summarizes the most relevant beam parameters for  the MDI design.
  \begin{table}[h!t]
	\setlength\tabcolsep{2pt}
	\centering
	\caption{FCC-ee beam parameters  Most relevant  for the IR design.}
	\label{tab:table1}
	\begin{tabular}{lccccc}
		\midrule
		Parameter  & Z  & $\rm W^{-}W^{+} $   &    ZH   &  $t\bar{t}$ \\
		\midrule
		$\rm E_{beam}$(GeV)   & 45.6  &  80    &     120    &  182.5\\
 Luminosity ($10^{34} {\rm cm}^{-2} {\rm s}^{-1}$)  & 230 & 28 & 8.5 &  1.55 \\
 Beam current (mA) & 1390 & 147 & 29 &  5.4 \\
 Particles/bunch ($10^{11}$)     &  1.7  &  1.5   & 1.8 &  2.3 \\
 Horiz. emittance (nm)           &  0.27  &  0.84  &  0.63  &   1.46 \\
 Vert. emittance (pm)            &  1.0    &  1.7   &  1.3  &   2.9 \\
 $\beta^{*}_{x}$ (m)           &  0.15    &  0.2   &  0.3  &  1.0 \\
 $\beta^{*}_{y}$ (mm)           &  0.8   &  1.0   &  1.0  &  1.6 \\
  $\sigma^{*}_{x}$ ($\mu $m)       &  6.4    &  13   &  13.7  &  38.2 \\
 $\sigma^{*}_{y}$ (nm)                 &  28   &  41  &  36       &  68 \\
SR bunch length (mm)     & 3.5 & 3 & 3.15 & 1.97 \\
total bunch length (mm)     & 12.1 & 6 & 5.3 & 2.54 \\
RF Acceptance (\%)     &   1.9   & 2.3 &   2.3 &  3.36 \\
DA energy accept. (\%)      &   1.3 & 1.3 &  1.7 & -2.8/+2.4\\
Rad. Bhabha Lifetime (min) & 68 & 59 &   38 &  40 \\
Beamstr. Lifetime (min)    & > 200 & > 200 & 18 & 18\\ 
		\bottomrule   
	\end{tabular}
\end{table}

\begin{figure}[h!b]
   \begin{center}
\includegraphics[width=.45\textwidth]{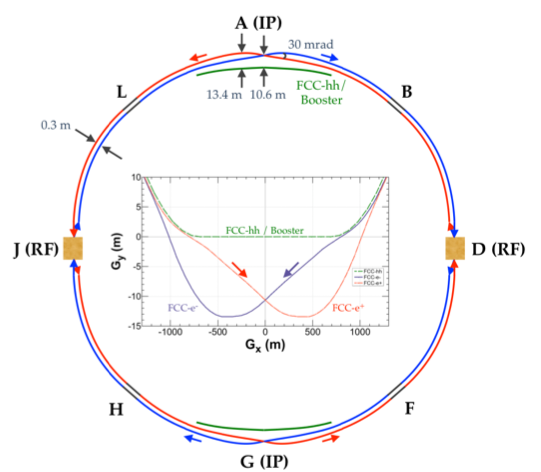}
   \end{center}
\caption{Schematic layout of the FCC-ee collider rings. The green line indicates the beamline of the FCC-ee booster  and  hadron collider FCC-hh. The plot in the middle shows the two beams trajectories at the IP.}
\label{fig:footp}
\end{figure}

\begin{figure}
   \begin{center}
\includegraphics[width=.45\textwidth]{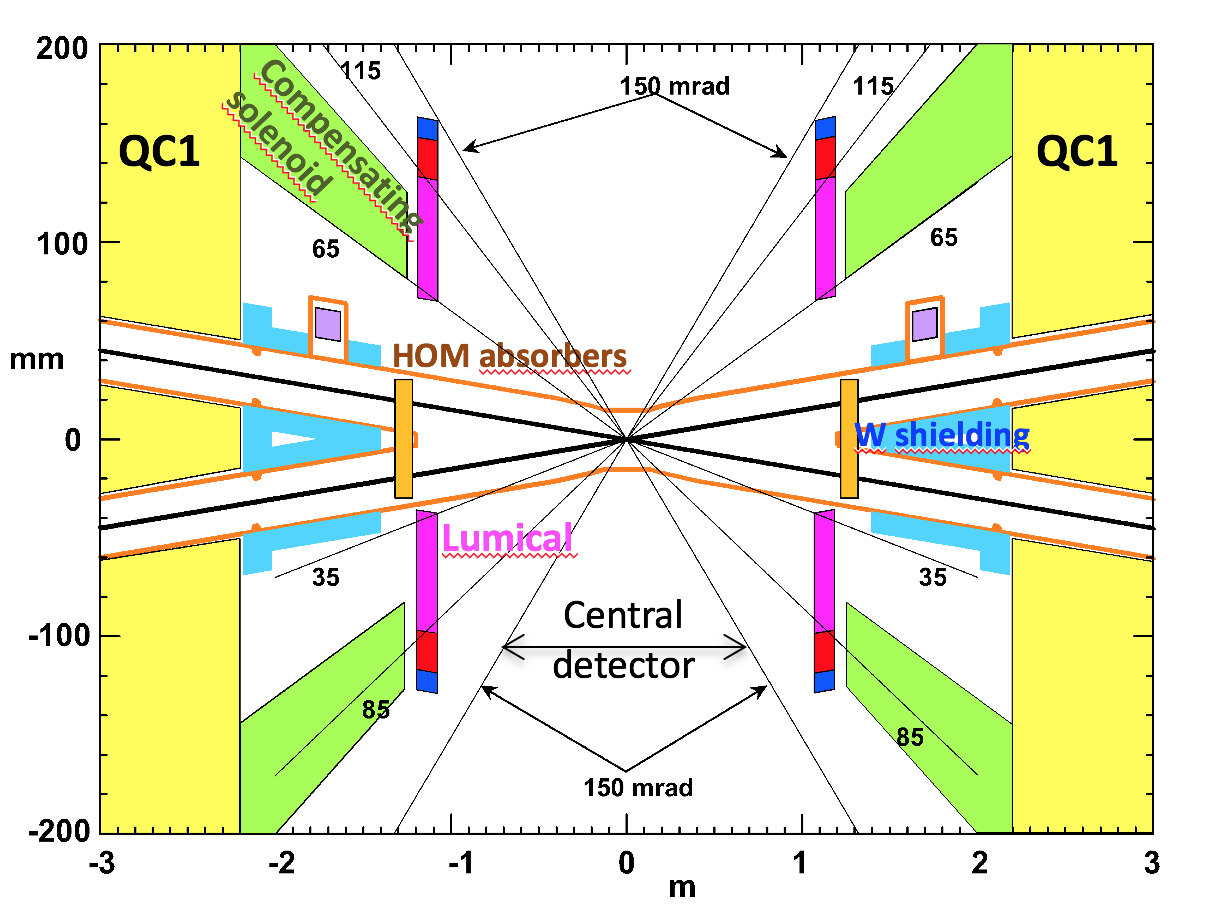}
   \end{center}
\caption{IR layout top view (x-z plane); note the expanded  scale for the ordinate ($\pm200$mm) with respect to the abscissa ($\pm$3m).}
\label{fig:IRlayout}
\end{figure}
To reach the target   luminosity  of $\rm 2.3\times10^{36}cm^{-2}s^{-1}$ at the Z-pole the crab-waist collision scheme is a necessary ingredient together with pushing the beam current to the limit, obtainable with double rings.
The baseline optics for the FCC-ee double-ring collider is described in Ref.~\cite{ref:prab-oide}.
The main characteristics of the optics design are two interaction points (IPs) per ring, horizontal crossing angle of
30~mrad at the IP and the crab-waist scheme with local chromatic correction system. A so-called tapering of the 
magnets scales all the magnetic fields with the local beam energy as determined by the SR. 
This optics is being improved and modified, for instance one of the most relevant modification for the IR design is the 
reduction of  $\beta_{x}^{*}$ to 15~cm at the Z to mitigate the coherent beam-beam instability~\cite{Oide:2017vvm}. 
Nominal emittances are very small and especially  in the vertical plane the  target value of $\epsilon_{y}=1~pm$ at the Z-pole
poses stringent requirements on  misalignment tolerances as well as on coupling correction.
 The design restricts the total synchrotron radiation (SR) power at 100~MW, thus the stored current per beam 
varies from 1.4~A at Z to 5.4~mA at $t\bar{t}$.
Following the LEP2 experience where the highest local critical energy was 72~keV for photons emitted 260 m from the IP~\cite{gvh}
 the FCC-ee optics design maintains critical energies from bending magnets below 100~keV
starting from 100~m from the IP; critical energy from the first bend after the IP 
is higher, being 691~keV at  $t\bar{t}$. 
 An asymmetric optics has been designed to meet these goals on the critical energy.
 The asymmetry allows each beam to come from the inner ring to the IP, to be bent strongly after the IP and to be merged back close to the opposite ring. 
 Outside the IR, the FCC-ee and FCC-hh trajectories are on the same footprint while  an additional tunnel is necessary for 1.2~km around the IP in order to allow for the crab-waist collision scheme with large crossing angle.
 The collider layout is shown in Figure~\ref{fig:footp} with  the two beam trajectories.  \par
Figure~\ref{fig:IRlayout} shows an expanded horizontal view for the region $\pm{3\,{\rm m}}$ from the IP.
The free length  between the interaction point (IP) and the first final focus quadrupole (QC1) $\rm L^{*}$ is 2.2~m.
The IR is symmetric and the two beam pipes  are  merged together  at about 1~m from the IP
and the distance between the magnetic centres of the two QC1 for the two beams is only few cm. 
In Figure~\ref{fig:IRlayout} are also shown the main components such as  the first focusing quadrupole named QC1 in yellow.
The first element at about 1~m from the IP is the luminosity counter, magenta in the plot and 
in red and blue the instrumentation and cables, followed by the compensating solenoid in light green and by the screening solenoid starting at about 2~m and out of this plot. 
The High Order Mode (HOM) absorber is in dark yellow,
the Tungsten shielding outside the vacuum pipe is in light blue. 
The detector solenoid, a cylinder with half-length 4~m and a diameter of around 3.8~m, is outside this picture. Its peak value is 2~T.
To reduce multiple scattering effects in the  luminosity monitor the vacuum chamber from $\pm{0.9\,{\rm m}}$ from the IP  
will be made of Beryllium followed by a Copper vacuum chamber throughout the final focus doublet. 
Synchrotron radiation mask tips  are also shown in the plot, they are placed in the horizontal plane just in front of QC1 at  2.1~m  from the IP to intercept SR scattered particles. At the mask tips the horizontal aperture will be reduced from 15~mm to 12~mm.
  The beam pipe will be at room temperature and water cooling is foreseen throughout  the IR, including inside the superconducting final focus quadrupoles.
The  compactness of the MDI design, as result of the available space to host all the necessary components  is a challenge.
An integration study with an assembly concept is in progress, to study the feasibility of this design.
In addition to the elements shown in Figure~\ref{fig:IRlayout}, also 
beam position monitors, flanges, bellows, cryostat, vacuum pump need to be placed.

\section{Solenoid compensation scheme}
The crab-waist collision scheme requires very small vertical beam sizes at the IP which implies the first final focus quadrupole
to be strong and close to the IP. It needs to be so close to the IP that it is located  inside the main detector solenoid. 
The additional ingredient for the crab-waist scheme is the large crossing angle, which brings the two beam trajectories to pass off-axis from the detector solenoid, inducing also an increase of the vertical emittance. To handle this unwanted effect, the detector solenoid 
field maximum is set at 2~T and, on the other hand,
a compensation solenoid is foreseen as close as possible to the IP. 
A screening solenoid is also needed to surround QC1 to avoid transverse beam coupling.
A  compact design is needed to leave a large physics detector acceptance; 
the accelerator components are required to stay  below an angle of 100~mrad from the beam axis.
This design gives an overall emittance blow-up estimate of 0.3~pm for two IPs at the Z-pole.\par
 The stringent requirements of the final focus quadrupoles can be satisfied by using canted-cosine-theta (CCT)  technology. It is an iron-free design with crosstalk and edge effect compensation, giving a field quality of better than one unit for all multipoles. Dipole and skew quadrupole correctors can be incorporated without increasing the length of the magnetic system.
The vertical emittance blow-up due to residual magnetic field has been estimated with SAD, finding an increase of 0.3~pm.
A full mechanical and thermal engineering analysis has been performed as well.
In order to prove feasibility, a prototype of this design is under construction at CERN.

\section{IR trapped modes and High Order Mode absorber}
\begin{figure}[h!b]
   \begin{center}
\includegraphics[width=.4\textwidth]{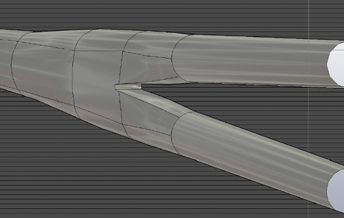}
  \end{center}
\caption{3D CAD view of the IR vacuum chamber in  the region where two beam pipes are merged together.}
\label{fig:cad_IR}
\end{figure}

The high beam currents produce electromagnetic waves in the IR.
The geometry of the beam pipe in the IR is as constant and smooth as possible to avoid unwanted electromagnetic trapped modes and heating problems. However,  in the IR the two beams
may generate electromagnetic waves where the vacuum chambers are combined into one 
near the collision point~\cite{Novokhatski:2017kpw}. This region is shown in  Figure~\ref{fig:cad_IR}.
\begin{figure}[h!b]
   \begin{center}
\includegraphics[width=.5\textwidth]{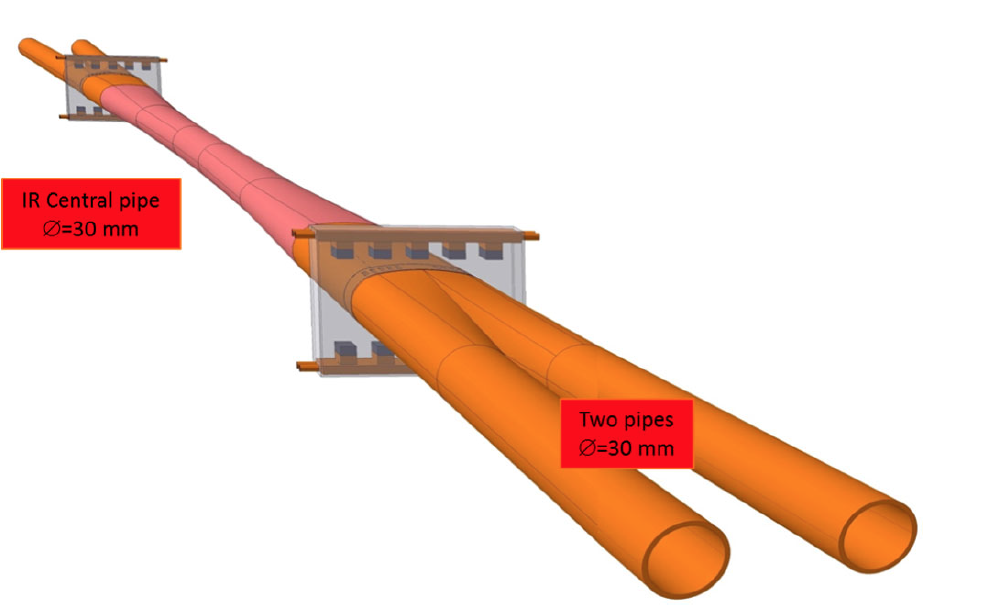}
  \end{center}
\caption{HOM absorbers design.}
\label{fig:HOM}
\end{figure}
The high order modes (HOM) that interact with a beam particle may 
cause local heating in the IR and typically their frequencies are in the range of several GHz.
Other electromagnetic waves, excited by the beam, with a frequency above the cutoff will travel
away from the IR and may cause heating downstream the ring. 
3D calculations have been carried out using CST~\cite{cst}  and HFSS~\cite{hfss} codes.
The numerical simulations show that there is a trapped mode with a frequency of 3.459~GHz 
and 2.91~kW power. This effect can be mitigated by
HOM absorbers, with slots oriented perpendicular to the HOM electric field, allowing the mode
field to easily propagate through these slots and, at the same time, the beam field, will not pass.
Water cooled absorbers are above and below the slots. 
They will be placed in the region where the two beam pipes are split in two, and just after the luminosity 
calorimeter, symmetrically from the IP.  
Figure~\ref{fig:HOM} shows the HOM absorbers design, which includes a water cooling system  to avoid heating.

\section{\label{sec:2}Synchrotron radiation}
The large crossing angle  
together with the high beam energy may induce  high  SR  in the IR and 
consequently into the detector.
We can state that the SR in the  IR drives the layout design. 
One of the most significant constraints  is the requirement on the                                                
critical energy and power of the synchrotron radiation generated upstream of the IR that  may shine into the detector. 
An additional constraint of the FCC-ee layout is the compatibility 
 with  FCC-hh, which drives the infrastructure design. 
In order to combine the two requests of a large 
crossing angle and the need to prevent high energy SR fans from going into the IP, the IR
                   optics have been designed asymmetrically so that the incoming beam from both sides comes from the
                   inner ring and the outgoing beam exits to the outer ring. In this way the outgoing beams are more
                   strongly bent than the incoming beams thereby lowering the SR energy from the incoming beams.
Independent approaches are used to evaluate  the main source of the SR background in the IR region
coming from photons emitted by beam particles passing through the last bending magnets and  by higher amplitude particles in the final focus quadrupoles.
\begin{table}  
\caption{Summary table of the SR coming from the last soft bend upstream the IP. Second column gives the incident number of photons  in the central beam pipe  per second.}
\begin{center}
\begin{tabular}{ccc}\hline
$\rm E_{beam}$  & $\rm E_{critical}$  &    $\gamma$  rate  on   \\ 
     GeV                 &   keV          &    central pipe  (Hz)        \\ \hline
  182.5             	  & 113.4.   	      &     $\rm 1.18\times 10^{8}$ \\
  175               		 &  100     	&    $\rm 1.25\times 10^{8}$       \\
  125                      &  36.4    	&   $\rm 1.01\times 10^{7}$   \\
  80                       &  9.56   	      &    $\rm 7.02\times 10^{5}$  \\
  45.6                  &  1.77     	      &    $\rm 9.58\times 10^{3}$  \\\hline
 \end{tabular}
\end{center}
\label{tab:sr}
\end{table}

 \begin{figure}[b]
   \begin{center}
\includegraphics[width=.42\textwidth]{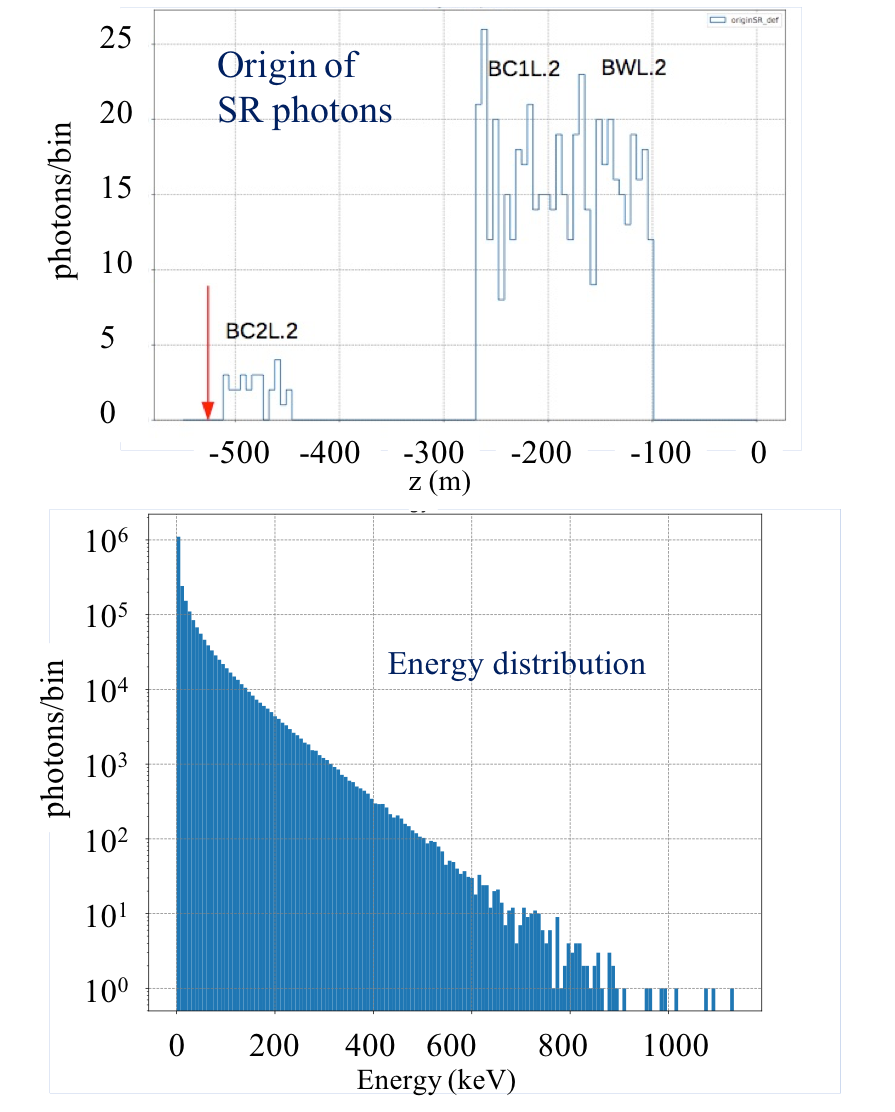}
  \end{center}
\caption{Upper plot: MDISim simulation showing the origin of the photons  generated by a beam starting at about -550~m from the IP (see the red arrow); the IP is at z = 0 m. Lower plot: energy distribution from the SR produced by the last bend upstream the IP.}
\label{fig:photon_origin}
\end{figure}
MDISim, $\rm SYNC\_BKG$ and SYNRAD+ are used to evaluate  the radiation critical energy, the SR fans and to design the IR layout including  masks, shieldings and the beam pipe.
MDISim~\cite{ref:mdisim} is a toolkit that combines existing standard tools MAD-X~\cite{madx}, ROOT~\cite{cernroot} and 
GEANT4~\cite{geant4, Allison:2016lfl}. It reads the MAD-X optics files, and uses its twiss output file to generate the geometry and the magnetic field information in a format which can be directly imported in GEANT4 to perform full tracking, including the generation of secondaries and detailed modelling of the relevant processes.
$\rm SYNC\_BKG$ traces beam macroparticles through sliced magnets and is a modified version
of a code developed at LBNL by Al Clark (see ref.~\cite{ref:prab_mdi} for a detailed description of the two methods and studies). 
SYNRAD+~\cite{synrad} is used to perform a full simulation of the optical interaction, including reflection as well as absorption, of the incident radiation with the beam pipe material. It takes as input a geometry either from CAD or STL format and the magnetic fields and it generates and tracks SR photons starting from a given beam distribution.

\begin{figure}[h!]
   \begin{center}
\includegraphics[width=.5\textwidth]{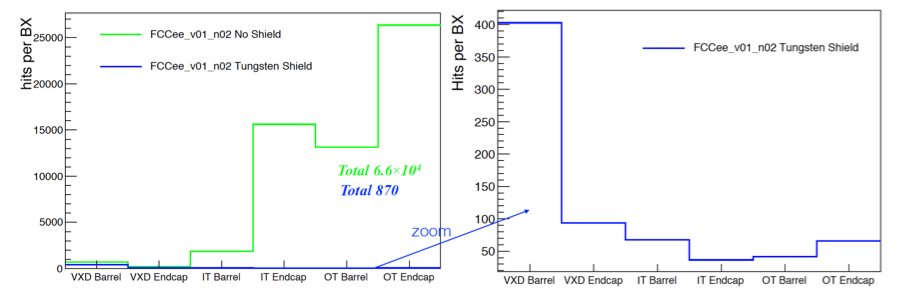}
  \end{center}
\caption{Hits per bunch crossing in the different CLD subdetectors (see the abscissa) due to SR with (blue line) and without (green line) Tungsten shielding.}
\label{fig:SR_W}
\end{figure}

To reduce SR backgrounds to tolerable limits the first criteria was to set a minimum distance for the bending magnets 
from the IP and the maximum critical energy for incoming beam.
The synchrotron radiation flux reaching the detectors can be further reduced by the combination of fixed and movable masks (collimators), as well as by optimizing surface to reduce X-ray reflection.
We foresee fixed mask tips at 2.1~m upstream of the IP, just in front of the first final focus defocusing quadrupole, in order to intercept this radiation fan and prevent the photons from directly striking the central Be beam pipe. The next level of SR background then comes from photons that strike near the tip of these masks, forward scatter through the mask and then strike the central beam pipe. At the  top energy, most of these scattered photons will penetrate the Be beam pipe and then cause backgrounds in the detector. 
To reduce the effect of this SR source on the experiment we propose to add a thin layer (of the order of 5~$\mu m$) of  high-Z and high conductivity   material such as gold inside the Be beam pipe. This will also 
minimize beam pipe heating from image charge currents. 
Table~\ref{tab:sr} is a partial summary of the SR study  with details about the photon rate from the last soft bend upstream the  IP
 for all the running beam energies of the FCC-ee. In this study the beam has been considered on-axis. No SR from dipoles or  quadrupoles hits directly the central beam pipe. Figure~\ref{fig:photon_origin} shows the MDISim simulation for the SR generated in the last bending magnets before the IP for the \textit{t\_208\_nosol} optics for the top energy.The upper plot shows the origin of these photons in the last bending magnets, as generated by a beam starting at the red arrow. The lower plot shows the energy distribution of the SR generated by the last bend before the IP.
 These SR photons 
   have been tracked  into the CLD and IDEA detectors, 
 showing good agreement with the  $\rm SYNC\_BKG$ simulation and   showing the effectiveness of the masking system.
Figure~\ref{fig:SR_W} shows the hits per bunch crossing with and without the Tungsten shielding.\par
As a next step more detailed simulations will be performed on the SR, namely using more realistic optics including the solenoidal field,
including misalignments and realistic beam distributions that may be slightly off-axis through the final focus quadrupoles. In addition,
presently GEANT4 doesn't include SR reflection and this effect will be studied as well. SR collimators to intercept far bends are planned, as from the LEP experience.
 
\section{\label{sec:5}Beam induced and luminosity backgrounds}
The deleterious effects of the background  is a very important issue in the IR, detector and Machine Detector Interface designs.
Beam induced backgrounds are scattering processes leading to particles loss; inelastic beam-gas scattering has been simulated in the MDI region. Scattering of the stored electrons and positrons with thermal photons responsible for single beam particle losses at high energies  is also presently under study as from~\cite{thermalphotons}.
Luminosity backgrounds are produced at the collision point.
Beamstrahlung induced backgrounds have been studied in full simulation.
They generate backgrounds at the interaction point and are mostly forward directed leaving the IR.
Detailed background studies are in progress to design the MDI region with proper shieldings and collimators.
The impact of machine beam losses in the detector is being considered with full GEANT4 simulation for all the background sources. 
We briefly describe the study performed for these background sources.

 \subsection{Inelastic  beam-gas scattering}
A precise and effective methodology to perform a detailed study of beam-gas scattering especially in the IR is provided by MDISim.
Beam gas induced background has been studied in B-factories (PEP-II and KEKB) and in Super-B factories. SuperKEKB is now starting to benchmark simulation studies with real data~\cite{paladino}. \par
\begin{figure}[b]
\centering
\includegraphics[width=7.cm,angle=0]{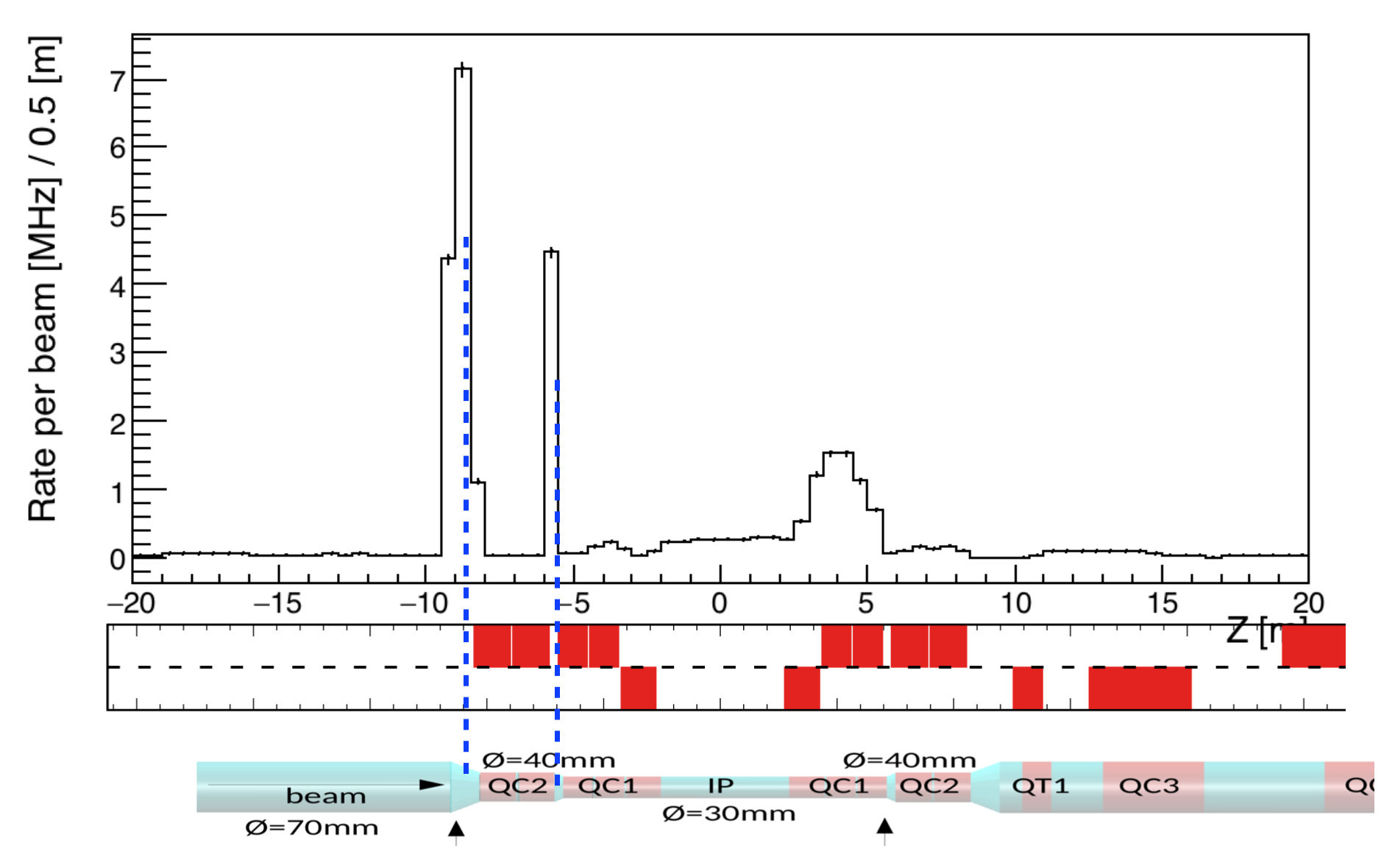}\caption{Loss map in the IR.The loss peaks correspond to the restriction of the vacuum chamber  between the last drift and final focus quadrupole QC2.}\label{beampipeIR}
\end{figure}

The MDISim toolkit was used to generate the files needed to perform a full GEANT4 simulation for 
the different beam running energies.
A constant beam pipe diameter of 70~mm is considered throughout the ring except for the section from -10~m to 10~m around the IP, shown in Figure~\ref{beampipeIR}. 
A constant gas pressure of $10^{-7}~Pa$ (or $10^{-9}~mbar$) is assumed for our study.

 \begin{table}[t]
\centering
\begin{tabular}{c|c|c|c|c}\hline 
 &I & R$_{MDI}$ &R$_{ZOOM}$ & R$_{MDI}$/I\\
& \small{[mA]} & \small{[MHz]} & \small{[MHz]} & \small{[MHz/A]}\\\hline\hline
Z & 1390 & $147$ & $29.2$ & $105$ \\
W & 147 & $15.8$ & $3.43$ & $107$ \\
H & 29 & $2.96$ & $0.536$ & $102$ \\
T & 5.4 & $0.526$ & $0.0959$ & $97$ \\\hline
\end{tabular}\caption{Expected particle loss rate both for 1~km of machine section ($R_{MDI}$), and for the $\pm$20~m around the IP~($R_{ZOOM}$), for all the four energy runs.}\label{RateTable}
\end{table}

 The vacuum chamber in QC2 has a  diameter of  40~mm, in QC1 of 30~mm. The transition is considered in the simulation with conical tapering from 30~mm to 40~mm as well as from 40~mm to 70~mm (from QC2 to the arcs) in about one meter of longitudinal distance.
The beam pipe with magnetic elements was reconstructed in the simulation from $\pm850$~m from the IP. Primary particles were generated starting at -830~m from the IP, with realistic distributions in transverse phase space, according to the optical parameters of the beam in that point of the machine, and tracked for about 1000~m.\par
The simulations were performed for a residual gas consisting of N$_2$ molecules.
This represents a worst case, since the actual residual gas 
in the beam pipe is expected to contain only a certain fraction of this molecule.
Table~\ref{RateTable} gives the expected particle loss rates both for the whole simulated machine section and for the $\pm$ 20~m around the IP,
 for all the four energy runs.
 We predict loss rates of roughly 100~MHz per Ampere of beam current around the IR. As expected,
the highest loss rate is found for the Z-pole energy, essentially due to  the high current configuration.
IR losses  are concentrated in the regions where the vacuum chamber gets smaller as the beam approaches the interaction point.\par
Beam-gas simulation results have been also weighted with a realistic  pressure profile evaluated with MolFlow~\cite{molflow}
 for the \textit{fcc\_213} optics at the top energy for about 600~m upstream the IP.
About 40\% of increase in the expected losses has been found. \par
  First estimates of the background induced  by these off-energy scattered particles in the luminosity calorimeter
  show that the  impact is at safe values, mostly thanks to the high-Z shielding~\cite{mogens}.

\subsection{$e^{+}e^{-}$ pairs and $\gamma \gamma \to$ hadrons}
 Beamstrahlung induced backgrounds have been simulated with GuineaPig++~\cite{guineapig}
namely coherent and incoherent pair creation (CPC and IPC) and $\gamma \gamma \to$ hadrons.
This effect  has been simulated through the detector with full simulation studies~\cite{Voutsinas:2017eca}. 
The Coherent Pair Creation (CPC) is strongly focused on the forward direction and is negligible at FCC.
The Incoherent Pair Creation (IPC) is expected to be one of the main sources of backgrounds.
The impact of this background source has been evaluated for the two FCC-ee proposed detectors CLD (Clic Like Detector) and IDEA. 
The CLD detector has been derived from CLIC detector model and it has been optimized for FCC-ee experimental conditions. CLD has a 2~T main solenoid field, with vertex detector (VXD) with 3 double layers (barrel) and 4
discs (barrel) and a Silicon tracker. The forward region has a cone of 100~mrad reserved for accelerator use.\par
The IDEA detector has a vertex detector (MAPS), an ultra-light draft chamber PID (DCH) and 2~T field.\par
The CLD maximum occupancy per bunch crossing for the IPC and SR is low, being about $\sim4\cdot10^{-4}$ in the VXD for the top energy and  $\sim1.6\cdot10^{-4}$ in the tracker.
The drift chamber average occupancy results low as well, being $\sim$2.9~\% for the IPC and $\sim$0.2~\%
for the SR at the top. 
 
\subsection{Radiative Bhabha and Beamstrahlung loss map}
Radiative Bhaba and Beamstrahlung are luminosity background sources that can cause beam losses in the IR also due to multiturn effects.  GuineaPig++~\cite{guineapig} and BBBREM~\cite{Kleiss:1994wq}  are used for the radiative Bhabha scattering generator, then multiple turns particle tracking is performed with SAD~\cite{sad} to determine  the IR  loss maps.
All the beam energies have been considerd, from 45.6~GeV to  to the top energy. 
At 45.6~GeV, the radiative Bhabhas are all lost up to about 70~m downstream the first IP.
At 175~GeV, the radiative Bhabhas are lost  mainly in the first half of the ring, and high energy particles that get eventually lost     reach the second IP.  
These particles loss distributions are then tracked into the sub-detectors with a full GEANT4 simulation.\par
For the Beamstrahlung background, the beam-beam element was inserted at both IPs  and  tracking for a thousand turns with full lattice was considered. 
  Particle losses are mainly concentrated 5~m around the IP in the vertical plane and losses happen mainly in the first few turns.
The impact on detector performance is still under investigation although we are expecting it not to be a major issue. 

\section{Conclusions}
The FCC-ee Machine Detector Interface baseline conceptual design is ready. 
Many details have been studied and  work is in progress to develop more and more realistic simulations. 
The solenoid compensation scheme foresees a vertical emittance blow-up of 0.3~pm. Improvements and alternative designs are on-going.\par
Beam induced and luminosity backgrounds have been studied. Loss maps have been evaluated for inelastic beam-gas scattering, and also for radiative Bhabha and beamstrahlung processes similar study is in progress. 
Luminosity processes have been directly simulated in the detectors and incoherent pair creation is the dominant effect but always at safe values for both detector designs, CLD and IDEA.
At the present level of simulations we find that the SR masking results are very efficient for protecting the detectors and maximum occupancy is  under control and at safe values.

\section{Acknowledgments}
We thank the FCC-ee team who contributed to discussions related to the Interaction Region and Machine Detector Interface design.

\bibliographystyle{unsrt}

\end{document}